\documentclass[usegraphicx]{mn2e}
\usepackage{times}
\newlength{\colwidth}
\begin{document}

\title[Interstellar extinction towards open clusters and galactic structure]
{Interstellar extinction towards open clusters and galactic structure}
\author[Y. C. Joshi]{Y. C. Joshi\thanks{E-mail: ycjoshi@tifr.res.in}\\
Tata Institute of Fundamental Research, Homi Bhabha Road,
Mumbai -- 400 005, India}
\date{Accepted ---------.
Received ---------;
}  

\pagerange{\pageref{firstpage}--\pageref{lastpage}} \pubyear{2005}

\maketitle

\label{firstpage}

\begin{abstract} 
We study the distribution of interstellar matter near the galactic plane
on the basis of open star clusters and reviewed the correlation of extinction
with different physical parameters of the clusters to understand the galactic
structure. It is seen from the extinction towards open clusters that about
$90\%$ of the absorbing material lie
within $-5^\circ \le b \le 5^\circ$ of the galactic plane. The mean thickness
of the absorbing material, which is determined in terms of half-width value
$\beta$, is estimated to be about $125\pm21$ pc. We show that the interstellar
absorption follows a sinusoidal variation with galactic longitude and
maximum and minimum absorptions occur at $l \sim 48^\circ\pm4^\circ$ and
$l \sim 228^\circ\pm4^\circ$ respectively. It is found that the galactic
plane defined by the reddening material is inclined by an angle of
$0^\circ.6\pm0^\circ.4$ to the formal galactic plane and inclination is
maximum
at $l \sim 54^\circ\pm6^\circ$. The reddening analysis has been used to
constrain the Solar offset which is found to be about $22.8\pm3.3$ pc above
the reddening plane. We obtained a scale height of $53\pm5$ pc for the
distribution of open clusters while it is $186\pm25$ pc for the distribution
of reddening material from the reddening plane. 
\end{abstract}


\section{Introduction}
Since long time, distribution of interstellar material has been used to
understand the galactic structure. A number of studies have been carried
out on the basis of reddening variation of stars and various parameters
like reddening plane, Solar offset, distribution of interstellar material
in the sky etc. were derived using different kind of objects. Fernie
(1968) studied the galactic structure on the basis of classical Cepheids.
FitzGerald (1968) analysed the interstellar extinction within a few kpc
of the Sun using colour excesses of 7835 O to M stars and found that the
reddening material is highly concentrated to the galactic plane. Neckel \&
Klare (1980) derived extinctions and distances for more than
11000 O to F stars and investigated the spatial distribution of interstellar
extinction in the direction of $|b| \le 7^\circ.6$.
Lately, open star clusters have been used to study the distribution of
interstellar extinction at low-galactic latitude rather than the stars.
This kind of study has an advantage over the stellar
study since the distances and reddening derived through open clusters are
regarded as more reliable than those derived through individual stars.
Furthermore the open star clusters span a wide range of galactic longitudes,
latitudes and distances from the Sun. Janes \& Adler (1982) studied the
galactic structure and distribution of clusters in the galactic plane from
a catalog of 434 open clusters. Pandey \& Mahra (1987, hereafter PM87)
used photometric data of the 462 open
clusters within 1.5 kpc of the Sun to investigate the spatial
distribution of interstellar extinction and concluded that the galactic plane
defined by the reddening material is inclined by an angle of
$0^\circ.8\pm0^\circ.2$ and the Sun is situated at a distance of about 10 pc
above the plane of symmetry defined by the interstellar matter. Pandey
et al.~(1988), on the basis of young open clusters of the same
sample, found
these values as $0^\circ.5\pm0^\circ.4$ and $28\pm5$ pc respectively. From the
same data, though, Janes et al. (1988) concluded that there are no
significant undulations around the plane of symmetry within 4 kpc of the Sun.

However, conclusions drawn from the photometric data of the open clusters
may have been affected by the considerable uncertainty in
their parameter determinations like distance, reddening, age etc. as well as
lack of homogeneous data.  During last ten years, new data on the open
clusters along with more precise determination of physical parameters have
become available
mainly due to large size telescopes and improved detectors like CCD which
allowed us to obtain deeper photometry of the clusters. Chen et al.~(1998)
derived an analytical expression for the interstellar extinction as a function
of galactic longitude and distance based on the observed distances and colour
excesses of 434 open clusters within $|b| \le 10^\circ$.
Recently Dias et al.~(2002) provided a comprehensive catalogue of more than
1600 open clusters which also include the information available in the
Lyng\.{a} catalogue (1987) as well as WEBDA catalogue (Mermilloid, 1995). The
catalogue contains almost all the clusters known till date.
It therefore seems worthwhile to re-investigate the classical problem of
galactic structure with this improved informations. In the present paper,
we investigate the spatial distribution of interstellar material and study the
Galactic structure using the physical parameters derived for a large
sample of open clusters.
The detail about the data used is explained in Sect. 2. In Sect. 3, we analyse
the spatial distribution of interstellar extinction while the Sect. 4 deals
with the correlation of interstellar extinction with the different physical
parameters of the clusters in order to understand
galactic structure. The main results of the present study are summarized
in Sect. 5.
\begin{figure}
\includegraphics[width=8.7cm,height=9.0cm]{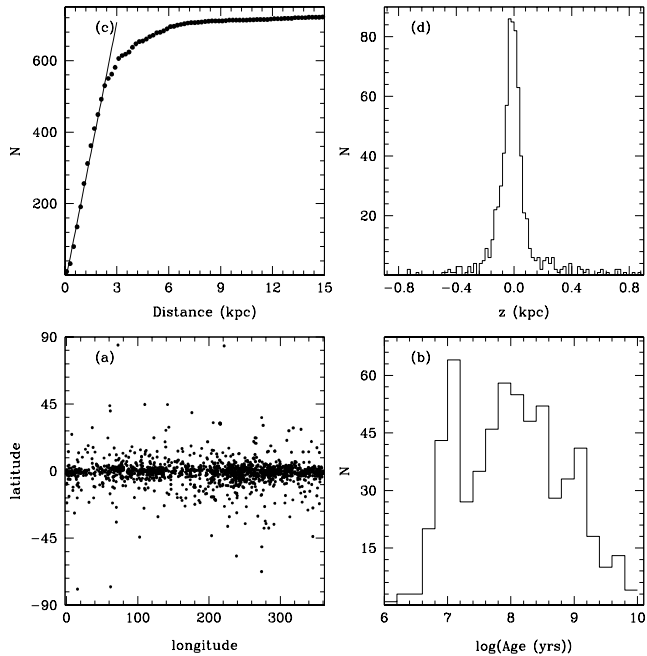}
\caption[]{
The distribution of open clusters in the $l-b$ plane is shown in (a). A
number distribution of the clusters in log(Age), distance from the Sun
(cumulative number) and the distance perpendicular to the galactic plane
are shown in (b), (c) and (d) respectively. Few odd clusters lying quite
far-away in the x-axis are not shown in (d) for the clarity of the histogram.
}
\end{figure} 
\section{The data}
Data for the present work is taken from the data base of open star clusters
compiled by W. S. Dias (Dias et al.~2002) at the web site
http://www.astro.iag.usp.br/$\sim$wilton/. Dias et al.~(2002) reported a
catalogue of 1599 clusters at the CDS, however, a total of 1632 clusters are
available in the online catalogue updated till March, 2005 in
which 747 clusters have distance and reddening informations. Out of them, we
have included only 722 clusters for our analysis since 20 of them are 
considered to be doubtful by the DSS images inspection (Dias et al.~2002)
while 5 are POCR (Possible Open Cluster Remnant, Bica et al.~2001). The age
informations of 602 clusters are also available out of these 722
clusters. The clusters are almost uniformly distributed in the longitude
$l$ but are densely packed in the latitude $b$ near the galactic plane as
seen in Fig.~1(a).
The age histogram drawn in Fig. 1(b) shows that the clusters are known in all
the age range from 6 Myrs to as much as 10 Gyrs. Though the clusters are
observed up to a distance of about 15 kpc, however,
it is seen that the number of clusters decreases with distance which suggests
of possible distance and/or reddening effect on the detectability of clusters.
When we plot the cumulative distribution of the clusters as a function of
distance from the Sun (see Fig. 1(c)), we found that the present cluster
sample is complete only up to a distance of about 2 to 3 kpc. It is seen from
the histogram of the
distance perpendicular to the galactic plane ($z = r \sin b$) plotted in Fig
1(d) that almost 90\% clusters lie within 200 pc of the galactic plane. We
notice that the clusters distribution is maximum at $z \sim -25$ pc which
indicates that the maximum reddening material may be lying below the galactic
plane. A further study of this effect shall be discussed in Sect. 4.2.
\begin{figure}
\includegraphics[width=9.0cm,height=9.0cm]{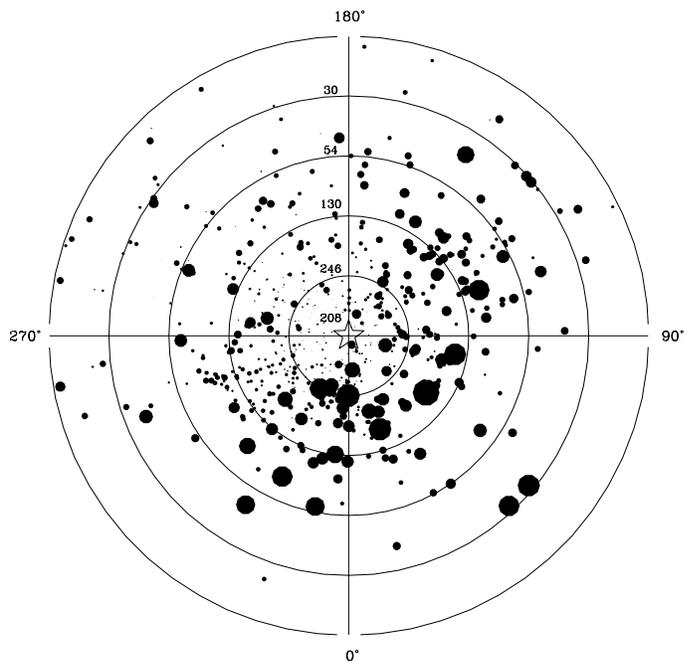}
\caption[]{
The distribution of open clusters as projected on the galactic plane. Points
are normalized to the extinction values such that the smallest and largest
points denote $E(B-V) = 0.01$ mag and 2.55 mag respectively. The position of
Sun is shown by a symbol of star at the center. Five concentric circles at
an equal distance of 1 kpc are drawn. The number of clusters in each distance
bin are also given.
}
\end{figure} 
\section{Interstellar extinction towards open clusters}
One of the characteristics of the interstellar extinction is its irregular
structure in the sky. Using the reddening of 722 open clusters which have
been determined independently through the analysis of colour and magnitude
of stars in the clusters, we map the extinction
variation in the sky that is shown in Fig.~2. All the clusters are put on the
galactic plane to draw the map. Since most of the clusters having reddening
information are observed within 5 kpc of the Sun, we map here
only first 5 kpc regions shown by 5 concentric circles at an equal distance
of 1 kpc around the Sun. The size of the dots are an indication of extinction
in the sense that bigger the dots, larger the extinction is. The map
shows a complex structure of the sky. One of the striking feature of the map
is clumpy behaviour of the cluster distribution at some places like
$l \sim 320^\circ - 350^\circ$. The concentration of the reddening material
to the galactic plane is quite visible in the region
$l \sim 330^\circ$  .. $0^\circ$ .. $150^\circ$ while the least absorption
lie in the region $l \sim 180^\circ$ to $270^\circ$. A galactic dust
distribution map up to 3 kpc distance from the Sun shown by Neckel \& Klare
(1980) also show similar picture.
\begin{figure}
\includegraphics[width=8.7cm,height=10.0cm]{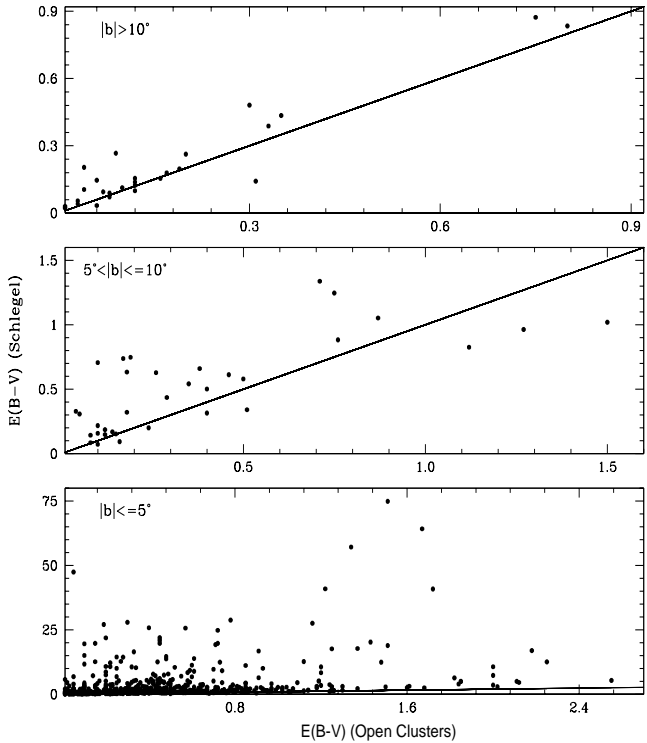}
\caption[]{
A comparison of the interstellar extinction towards opon clusters with the
Schlegel map for 3 different latitude range. The dashed lines denote the 
y = x line.
}
\end{figure} 

We compared the $E(B-V)$ values for the open clusters given in the Dias
et al. (2002) with the extinction map given by Schlegel et al. (1998,
hereafter SFD98) which is based on the IRAS 100 $\mu$m surface brightness
converted to extinction and a comparison in three different latitude bins are
shown in Fig.~3. We found a large discrepancy in the extinctions values towards
open clusters in the SFD98 reddening map below $b \le 5^\circ$. It is not
unexpected since most contaminating sources have not been removed from the
SFD98 maps as well as temperature structure of the Galaxy is not well resolved
at low galactic latitude. The comparisons at $b > 5^\circ$ show that the
SFD98 map is overestimated by a factor of $1.04\pm0.09$ and $1.12\pm0.05$ in
comparison to the extinction based on the open clusters in the regions
$5^\circ < b \le 10^\circ$ and $b > 10^\circ$ respectively. Chen et al. (1999)
also found that SFD98 reddening map overestimates $E(B-V)$ by a factor of 1.16
and 1.18 based on the study of globular and open clusters respectively at
$|b| > 2.5^\circ$. In recent times, various
authors have compared interstellar extinction in different regions of the
sky with the SFD98 reddening map (cf. Arce \& Goodman, 1999, Cambr\'{e}sy et
al. 2001, 2005, Dutra et al. 2002, 2003a, 2003b) and supported the fact that
SFD98 map overestimates the extinction in many parts of the sky, if not in the
whole sky.
\begin{figure}
\includegraphics[width=8.7cm, height=10.0cm]{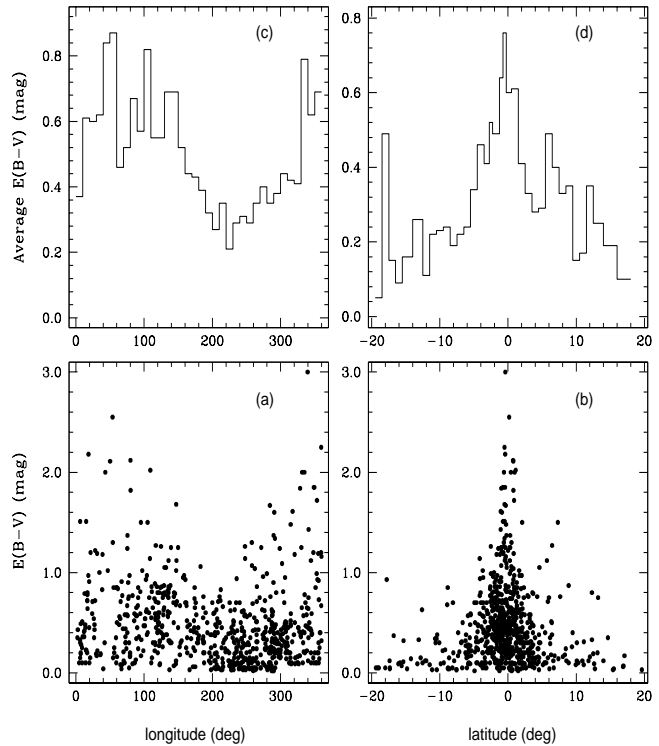}
\vspace{0.05cm}
\caption[]{
The variation of $E(B-V)$ vs $l$ (left) and $b$ (right). In the
upper panels, the histogram of the mean value of reddening using bin
sizes of $10^\circ$ in $l$ and $1^\circ$ in $b$ are shown.
}
\end{figure} 

In order to study the distribution of extinction along the galactic longitude
and latitude, we plot E(B-V) against $l$ and $b$ in Figs.~4(a) and 4(b); and
distribution of the mean value of $E(B-V)$ in the $10^\circ$ bin in $l$ and
$1^\circ$ bin in $b$ in Figs.~4(c) and 4(d) respectively, however, around the
peak in Fig.~4(d), we made smaller bin size of $0.5^\circ$ due to higher
concentration of clusters. We have calculated the mean value of reddening
($\overline{E(B-V)}$), concentration of reddening ($C_{red}$) which is
defined as $E(B-V)$ in the sub-region divided by total $E(B-V)$ of the
clusters and number of clusters ($N_{CL}$) for 5 different latitude range
which are given in Table 1. From the figures and table, one can easily
notice following approximate though important features:
\begin{enumerate}
\item That the  reddening is comparatively larger and scattered in the
direction of galactic center than the small and systematic variation in
anti-center direction (see Figs 4(a) and 4(c)). On an average, the reddening
is almost twice in the region $l \sim 330^\circ$ .. $0^\circ$ .. $150^\circ$
than the rest of the region.
\item Most of the open clusters are concentrated near the galactic plane.
However, we notice that the peak of the reddening is shifted by
$\sim 0.5^\circ$ in the southwards of the galactic plane (see Figs.~4(b)
and 4(d)). 
\item A simple approximation on the basis of $E(B-V)$ informations of the
open clusters indicate that about $90\%$ reddening material lie within
$b \pm 5^\circ$ of the galactic plane (see Table 1).
\end{enumerate}
\begin{table}
\centering
\caption{The distribution of the clusters and reddening in different latitude
range.}
\begin{tabular}{cccc}
\hline
range in $|b|$&$\overline{E(B-V)}$&$C_{red}$& $N_{CL}$\\
(deg)         &  (mag)         & (\%)           &      \\ \hline
0-2           &  0.59          &        67      &383   \\
2-5           &  0.40          &        22      &190   \\
5-10          &  0.29          &        7       &84    \\
10-20         &  0.19          &        3       &54    \\
$>$20         &  0.08          &        1       &11    \\
\hline
\end{tabular}
\end{table}
\begin{figure*}
\includegraphics[width=18cm,height=14.7cm]{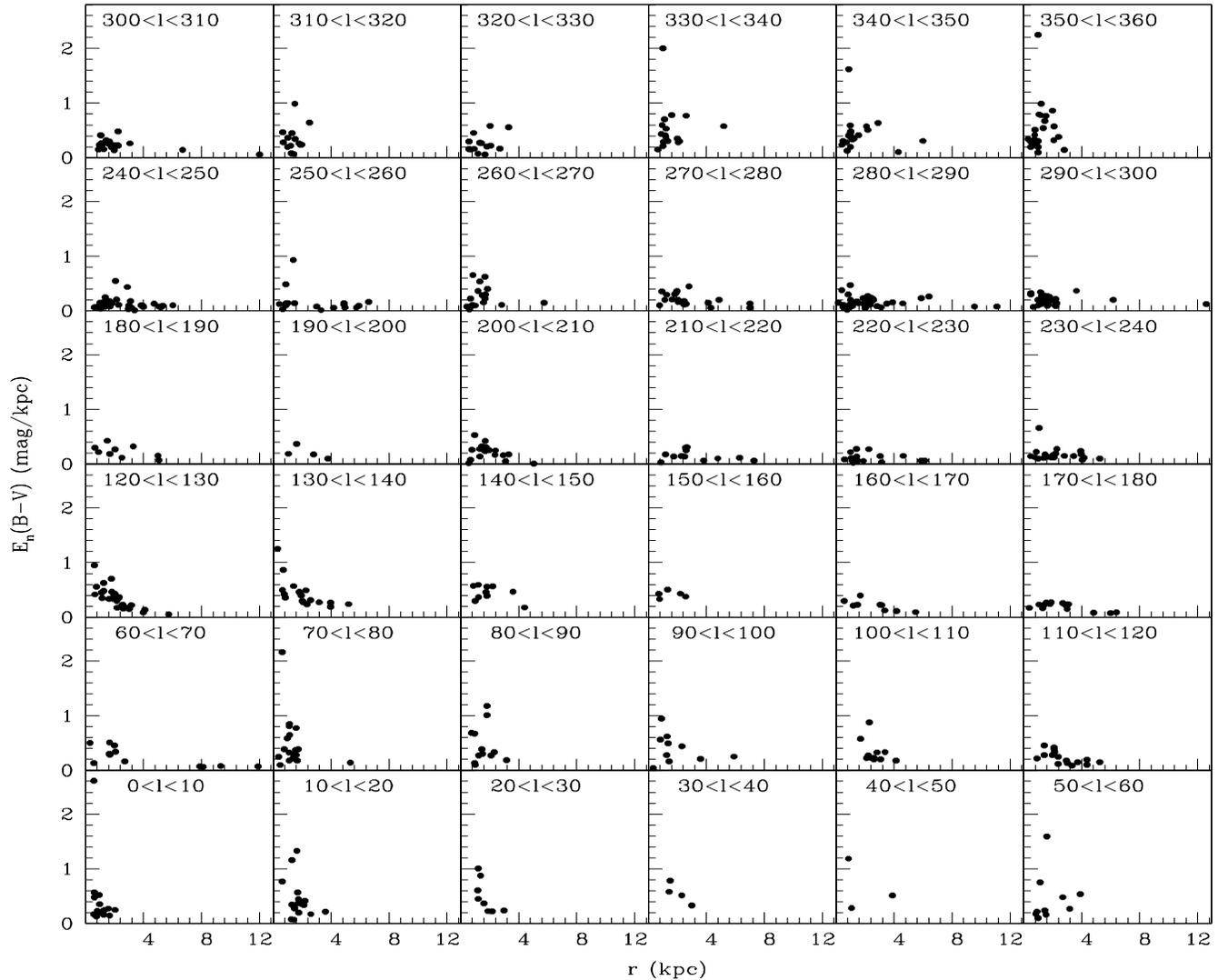}
\caption[]{
The normalized extinction $E_n(B-V)$ versus distance $r$ for 36 zones of
galactic longitude. Range of the longitudes for each zone are shown at the top
of each plot.
}
\end{figure*} 
It is seen from the age distribution of clusters under study that while
young open clusters are confined very close to the galactic plane and have a
clumpy distribution, the old age clusters are mostly distributed in the
far-off region from the galactic plane. Our study shows that the reddening
layer is thicker near the galactic plane and about 80\% of clusters sample
and about 90\% reddening material lie within
$-5^\circ \le b \le 5^\circ$ of the galactic plane. We, therefore, consider
only those clusters for the further analysis which are lying in this region
of the Galaxy.

\section{Analysis and Results}
\subsection{Variation of extinction with distance}
It is a well known fact that the interstellar matter is extremely inhomogeneous in the sky, therefore, we study the behaviour of extinction variation in the
sky by dividing it in small zones. In some previous studies of the galactic
structure which have been carried out on the basis of extinction variation in
the stars (e.g. Neckel \& Klare, 1980), the sky has been
split up in very small sub-regions. Chen et al. (1998), on the way to construct
extinction map in the galactic plane below $|b| \le 10^\circ$ divided the
cluster sample into 36 cells with $\Delta l = 10^\circ$. Owing to the
limitations of smaller sample of the clusters than the stars, we also divide
\begin{figure}
\includegraphics[width=8.5cm,height=10.5cm]{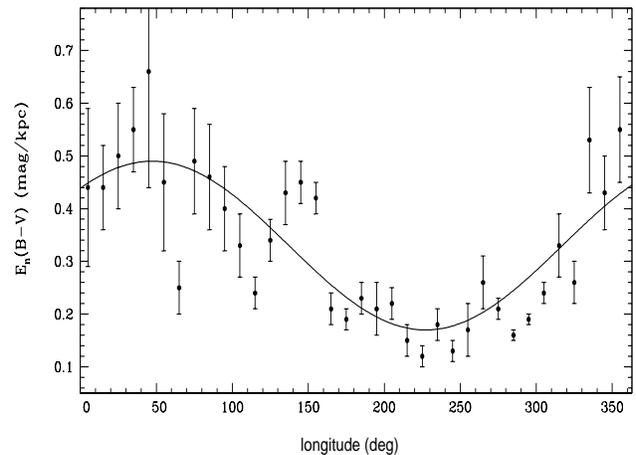}
\vspace{-4.5cm}
\caption[]{
The mean value of the normalized extinction as a function of $l$. A best fit
sinusoidal variation is shown by continuous line.
}
\end{figure} 
\begin{figure*}
\includegraphics[width=16cm,height=13cm]{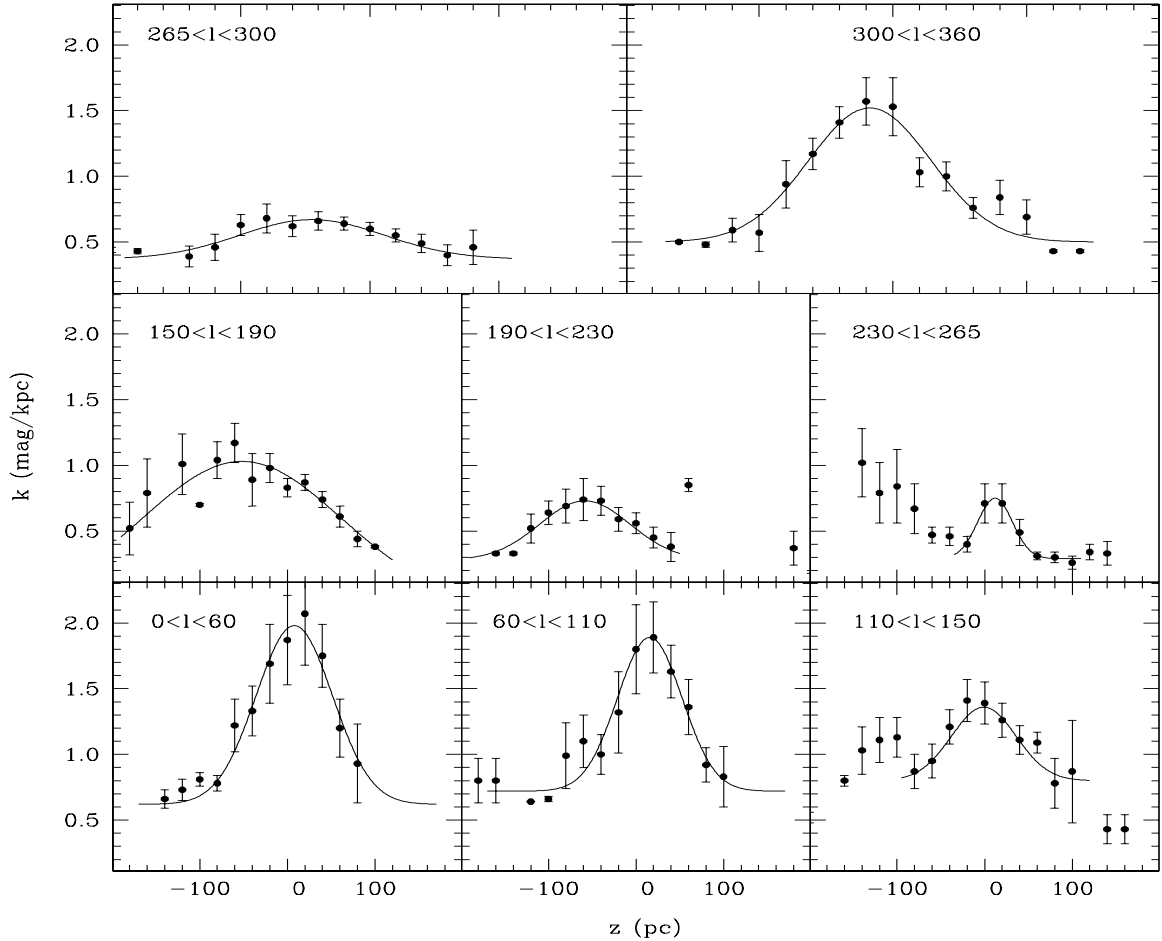}
\caption[]{
The variation of $k$ as a function of z for 8 different regions in longitude
which are written at the top of each diagram. A least square Gaussian fit around
the maximum absorption is also drawn.
}
\end{figure*} 
our sample of clusters located in $|b| \le 5^\circ$ into 36 zones of $10^\circ$
interval in longitude. We normalize the extinction by distance and a term,
$E_n(B-V)$, called normalized extinction has been evaluated as
$$E_n(B-V) = \frac{E(B-V)}{r}$$
where $r$ is the distance of cluster from the Sun. A variation of $E_n(B-V)$
with the distance from the Sun for all the 36 zones are drawn in Fig.~5. In
general, we see following features in the diagrams:
\begin{enumerate}
\item Within individual zones, the reddening shows a considerable but
irregular variation which cannot be approximated by any simple analytical
function.
\item The scattering in extinction variation is more in the direction of
galactic center where most of the complex dust clouds are believed to be
located.
\item The $E_n(B-V)$ changes significantly from one zone to another zone from
about 0.1 to 0.6. A similar variation has also been noticed by Chen et al.
(1998) near the galactic plane.
\item There is a general trend in the variation of $E_n(B-V)$ that it is
maximum around the longitude range $30^\circ - 50^\circ$ and minimum around the
longitude range $220^\circ - 250^\circ$. This kind of variations are also seen
by Arenou et al.  (1992) and Chen et al. (1998) in their extinction models.
\end{enumerate}
We determine the mean value of normalized extinction for each zone which is
plotted as a function of galactic longitude in Fig.~6. The variation can be
described as a sinusoidal function and a least square fit gives
$$ E_n(B-V) = 0.33 + 0.16 \sin(l + 43)$$  
\vspace{-0.75cm}
$$~~~~~~~~~~~~~~~~~~\pm0.01~\pm0.02~~~~~~~~\pm6$$  
where $E_n(B-V)$ is in mag/kpc and $l$ in degree. From the equation, it is estimated
that the extinction varies rapidly with distance in the direction of
$l \sim 50^\circ$ while it varies with a very slow rate in the direction
of $l \sim 230^\circ$. The mean extinction in the region of
$160^\circ<l<300^\circ$ is only $\sim 0.07$ mag/kpc which suggests that the
sky is relatively transparent in this direction of the Galaxy.

\subsection{Variation of absorption with z}
To study the distribution of interstellar extinction perpendicular to the
galactic plane, we initially determined the interstellar absorption $k$ as
$$ k = \frac{A_v}{r} = R\frac{E(B-V)}{r}$$
where $R$ is the total-to-selective absorption which we have taken as 3.1.
Therefore, above equation can be expressed as
$$ k = 3.1 E_n(B-V) $$
Since the different parts of the sky do not behave similarly, we therefore
divide the sky in 8 zones of the galactic longitude.
The boundaries of these zones are chosen in such a way that the
larger boundaries are made towards galactic center where smaller
number of clusters are seen per longitude bin due to higher extinction while
smaller boundaries are made towards galactic anti-center due to
opposite behaviour. Though it makes irregular boundaries but render lesser
scatter within the fields. The boundaries of the respective zones are
given at the top of each panel of Fig.~7 where we show the variation of
absorption in all the 8 zones
as a function of $z$. Here we have plotted the running mean of $k$ for
$60$ pc interval of $z$ against the mean value of $z$ of that interval.
A Gaussian fit around the central peak value of $z$ in all the diagrams
of Fig.~7 is also drawn. Although the distribution of interstellar
absorption varies significantly from one zone to another but the strong
concentration of absorption in the galactic
plane is clearly visible in all the zones except $230^\circ<l<300^\circ$.
Guarinos (1991) has pointed out that between Orion and Sco-Cen complex of the
above zone, the interstellar medium is transparent around the Sun up to at
least 1.0 to 1.4 kpc. Neckel \& Klare (1980) reported lowest extinction
in the region $210^\circ<l< 250^\circ$ from the analysis of early type stars
within 1 kpc of the Sun. Chen et al. (1999) also noticed deviation
in the region $225^\circ<l< 290^\circ$ in order to constrain the extinction
model based on the COBE/IRAS all sky reddening map.

The thickness of the absorbing layer in different zones can be expressed in
terms of half-width value, $\beta$, which is defined as the separation of $z$
values at the $1/e$ of the maximum value of interstellar absorption (Neckel,
1966). We determined thickness for the each zone and a final value of the
$\beta$ has been derived as a mean of the thickness which is estimated to be
$125\pm21$ pc in the $z$ direction. FitzGerald (1968) has reported $\beta$
in the range of
40 to 100 pc on the basis of colour excess versus distance diagram of about
8000 O to M type stars while studying open clusters in $|b|<10^\circ$, PM87
found it greater than 100 pc for most of the sub-regions of different
longitude range in the sky.
\begin{figure}
\includegraphics[width=8.5cm,height=10.5cm]{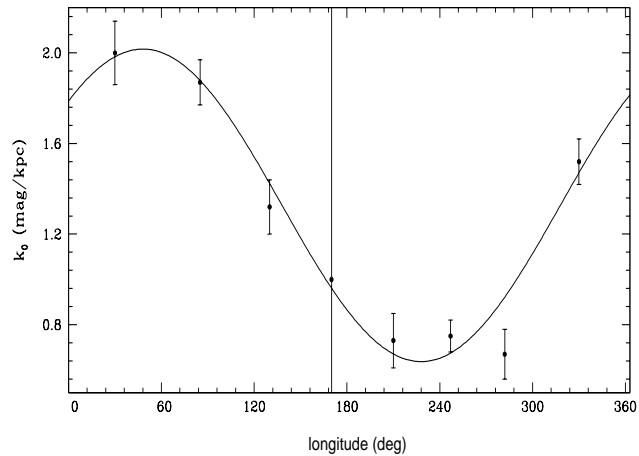}
\vspace{-4.5cm}
\caption[]{
Maximum absorption $k_0$ as a function of $l$.
A least square sinusoidal fit is drawn by a continuous line.
}
\end{figure} 

In order to analyse the absorption variation in more detail, we have 
determined the maximum value of absorption $k_0$ and corresponding distance
from the galactic plane $z_0$ from the fit. The variation of $k_0$ as a
function of mean value of longitude for each zone is drawn in Fig.~8 which
shows a sinusoidal kind of variation. A least square solution to the sinusoidal
function gives
$$k_0 = 1.30 + 0.69 \sin(l + 42)$$  
\vspace{-0.75cm}
$$~~~\pm0.03~\pm0.04~~~~~~~~~\pm4$$  
The absorption is found to be maximum in the direction of
$l \sim 48^\circ\pm4^\circ$ and it is minimum towards
$l \sim 228^\circ\pm4^\circ$. Fernie (1968) has also found a maximum average
absorption towards $l \sim 50^\circ$ based on the study of classical Cepheids.
Based on the study of open clusters, Lyng\.{a} (1982) and PM87 have found
$l \sim 230^\circ$ and $l \sim 50^\circ$ for the minimum and maximum
absorptions respectively. Using an inverse technique of Tarantola \& Valette
(1982) on open clusters, Chen et al.~(1998) found the lowest absorption in the
direction of $l = 210^\circ$ to $240^\circ$, however, they reported the highest
absorption towards $l = 20^\circ$ to $40^\circ$ which is slightly lower than
those found in all other studies. 

To study the variation of maximum absorption perpendicular to the galactic
plane, we plotted the value of $z_0$ against the mean value of longitude
in Fig.~9.  A least square solution for a sinusoidal function gives
$$z_0 = -22.8 + 41.1 \sin(l + 36)$$  
\vspace{-0.75cm}
$$~~~~~\pm3.3~\pm4.8~~~~~~~~~~~~\pm6$$  
We have excluded the lone point around $230^\circ<l<265^\circ$ in drawing the
fit where $z_0$ has quite a large shift from the sinusoidal function. From the
figure, we draw following conclusions:
\begin{figure}
\includegraphics[width=8.5cm,height=10.5cm]{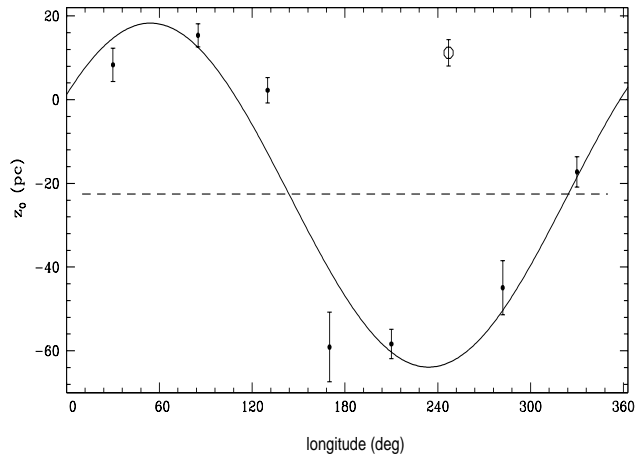}
\vspace{-4.5cm}
\caption[]{
The height z above or below the galactic plane at the maximum absorption $k_0$
as a function of $l$. A least square sinusoidal fit is shown by a continuous
line while dashed line shows the mean value. The lone point in the vicinity
of $l \sim 250^\circ$ drawn by open circle is omitted from the fit.
}
\end{figure} 
\begin{enumerate}
\item That the maximum extinction lies above the galactic plane towards
galactic center and below the galactic plane towards galactic anti-center.
\item It shows that the distance of the galactic plane at maximum
absorption is symmetric around $z \sim -22.8$ pc which indicates that the
larger amount of reddening material is lying below the galactic plane which
is in confirmation with our approximate estimation in the earlier
section.
\item A symmetric variation around $z \sim -22.8$ pc means that the galactic
plane of symmetry defined by the reddening material is lying below by the same
value to the formal galactic plane ($b = 0^\circ$ plane). This suggests that
the Sun probably lies
about $22.8\pm3.3$ pc above the galactic plane defined by the reddening
material. The offset of the reddening plane from the galactic plane has
important bearing on the determination of density distribution of different
kind of stars, particularly young stars, which are closely situated near the
galactic plane
(M\'{e}ndez \& van Altena 1998). Numerous studies have been carried out
to determine Solar offset from the reddening plane using different kind of
objects (see Table 2 for few of them). Most of the recent studies show a shift
in the
range of $15-30$ pc in the north direction of the galactic plane. Our
determination of Solar offset is thus in close agreement with these estimates.
\item The galactic plane seems to be tilted and maximum upward tilt is found
to be in the direction of $l \sim 54^\circ\pm6^\circ$. PM87 and Pandey et al.
(1988) have also reported an upward tilt in the direction of $l \sim 60^\circ$
and $50^\circ$ respectively.
\end{enumerate}
\begin{table}
\caption{The value of Solar offset above the reddening plane in previous
studies.}
\begin{tabular}{cll}
\hline
   $z_\odot (pc)$     &Based on the study           & Reference \\ \hline
$15\pm3$ &Wolf-Rayet stars                  & Conti \& Vacca (1990)\\
$15.5\pm0.7$ &IRAS point-source counts      & Martin (1995)\\
$20.5\pm3.5$ &optical star counts           & Humphreys \& Larsen (1995)\\
$\sim 14$ &near-IR surface brightness map   & Binney et al.~(1997)\\
$10-12$ &OB stars within $|b| \le 10^\circ$ & Reed (1997)\\
$27\pm3$ &star catalogue and simulation     & M\'{e}ndez \& van Altena (1998)\\
$27.5\pm6$&extinction model                 & Chen et al.~(1999) \\ 
$24.2\pm1.7$&O-B5 stars                     & Ma\'{i}z-Apell\'{a}niz (2001)\\ 
$22.8\pm3.3$&open clusters within $|b| \le 5^\circ$ & present study \\
\hline
\end{tabular}
\end{table}
\subsection{Reddening plane versus galactic plane}
It is a well known fact that the formal galactic plane does not coincide
with the planes of symmetry defined by the different objects like neutral
hydrogen layer (Gum et al., 1960), Cepheid variables (Fernie, 1968), WR stars
(Stenholm, 1975), HII regions and supernova remnants (Lockman, 1977), the
molecular clouds (Cohen \& Thaddeus, 1977, Magnani et al., 1985), Open
clusters (Lyng\.{a}, 1982, Pandey et al. 1988) etc. To examine the
relationship between reddening plane and formal galactic
plane, we also investigate the distribution of $z$ in a 1 kpc bin as a
function of $r$ following the approach given by Fernie (1968).
This we determined in two different directions having maximum
absorption i.e. towards $l$ = $54^\circ \pm 35^\circ$ and
$l$ = $234^\circ \pm 35^\circ$. A variation in the mean value
of $z$ as a function of distance is plotted in Fig.~10 where distances towards
$l$ = $54^\circ \pm 35^\circ$ are taken with positive sign while distances
towards $l$ = $234^\circ \pm 35^\circ$ are taken with negative sign.
If $z_\odot$ is the distance of the Sun from the reddening plane and $\phi$
is the angle between reddening plane and formal galactic plane then the
relation between two planes can be expressed as
$$z = r \sin\phi + z_\odot$$
where both $z$ and $r$ are expressed in pc. Since we have earlier determined
$z_\odot = 22.8\pm3.3$ pc, we put this value here and a least square regression
fit gives us
$$\phi = 0.6^\circ \pm 0.4^\circ$$
This shows that the reddening plane is inclined by an angle of
$0.6^\circ \pm 0.4^\circ$ with respect to the galactic plane. PM87 found an
angle of $0^\circ.8\pm0^\circ.2$ between these two planes while Pandey et
al.~(1988) reported it as $0^\circ.5\pm0^\circ.4$.
All the results are thus consistent within their
quoted errors. If we assume that the Sun lies at a distance of 8.5 kpc from the
galactic center then we estimate that the reddening plane may cut the formal
galactic plane at a distance of $6.3\pm0.9$ kpc from the galactic center.
\begin{figure}
\includegraphics[width=8.5cm,height=10.5cm]{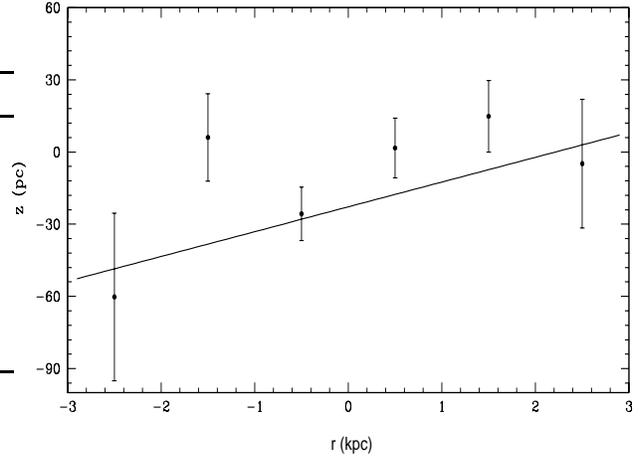}
\vspace{-4.5cm}
\caption[]{
Mean value of the $z$ in a 1 kpc bin as a
function of distance in the direction of $l = 54^\circ\pm35^\circ$ (taken as
a positive distance) and $l = 234^\circ\pm35^\circ$ (taken as a negative
distance). The least square fit is shown by a continuous line.
}
\end{figure} 
\subsection{Determination of scale heights}
To determine the scale height of the distribution of open clusters from the
reddening plane, we defined the plane of symmetry due to reddening material as
$$z^{\prime} = z_\odot + r \sin b \cos \phi - r \cos b \sin \phi cos (l-\theta_t)$$
where $\phi$ is the inclination angle between two planes and $\theta_t$ is the
angle of maximum inclination of the reddening plane. Other
symbols have their previous meanings. Suppose a large scale volume density of
the reddening material, n(r), follows a decaying exponential law with respect
to the reddening plane then it can be expressed as
$$n(z^\prime) = n_0 \exp(-|z^{\prime}|/h)$$
where $h$ is the scale height of the open cluster distribution from the
reddening plane. The mean value
of $z^\prime$ has been calculated in 20 pc intervals of $z^\prime$ and number
of open clusters are counted in each bin. In Fig.~11, we have plotted the
logarithm of number distribution of star clusters as a function of $z^{\prime}$.
From a least square linear fit, we determined
$$h = 53\pm5 ~ pc$$ 
The percentage of
clusters within $z^\prime$  of the reddening plane can be determined as
$$n(z^\prime) = \lbrack 1-exp(-|z^\prime|/h) \rbrack \times 100 \% $$
It shows that about 90\% of the open clusters are confined within a distance of
$122\pm12$ pc of the reddening plane.
\begin{figure}
\includegraphics[width=8.5cm,height=10.5cm]{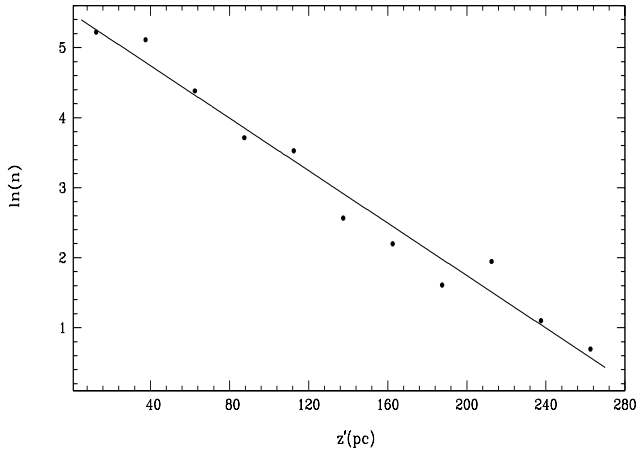}
\vspace{-4.5cm}
\caption[]{
The density distribution of clusters as a function of $z^\prime$. The vertical
axis is in the natural logarithm of $z^\prime$. Continuous line is
a least square fit to the points.
}
\end{figure} 
\begin{figure}
\includegraphics[width=8.5cm,height=10.5cm]{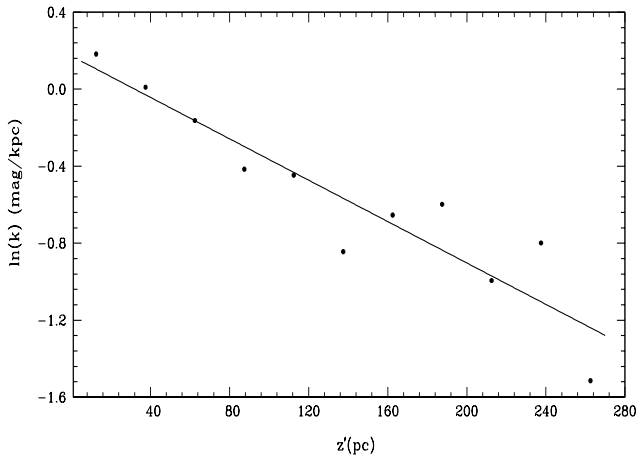}
\vspace{-4.5cm}
\caption[]{
The density distribution of absorption as a function of $z^\prime$. The vertical
axis is in the natural logarithm of $z^\prime$. Continuous line is
a least square fit to the points.
}
\end{figure} 

Similarly, to determine the scale height of the distribution of interstellar
material normal to the reddening plane, we draw logarithm of the mean value of
$k$ in 20 pc interval of $z^\prime$ as a function of $z^\prime$ in Fig.~12.
We further fit a linear regression line of the form
$$k(z^\prime) = k_0 \exp(-|z^{\prime}|/h^\prime) $$
where $h^\prime$ is the scale height of the interstellar material from
the reddening plane. A least square fit gives 
$$h^\prime = 186\pm25 ~ pc$$ 
From the above equation,
it is found that about 90\% of the reddening material lie within a distance
of $430\pm60$ pc from the reddening plane.

Pandey et al. (1988) estimated a scale height for the distribution of open clusters $h = 52\pm6$ pc from their study of
nearby open clusters while Lyng\.{a} (1985) reported a
scale height of 60 pc based on the study of young open clusters. Our value is,
thus, in agreement with the earlier estimates.
The scale height of the distribution of reddening material determined by PM87
as $160\pm20$ pc which is smaller than our result but their value is
estimated with respect to the formal galactic plane which is shifted by
a distance of $22.8\pm3.3$ upwards from the reddening plane.
\section{Summary}
In order to study the distribution of the interstellar extinction in the
sky and investigate the galactic structure in this context, we used the
data base of 722 open star clusters available at the website
http://www.astro.iag.usp.br/$\sim$wilton/ which is complied by the Dias
et al.~(2002).
On the basis of reddening of open clusters within $|b| \le 5^\circ$, we found
that the extinction varies with distance in a very slow rate of about 0.07
mag/kpc in the region of $160^\circ<l<280^\circ$, however, it is highly
variable in other parts of the Galaxy. We determined that almost 90\% of the
interstellar material lie within $5^\circ$ of the galactic plane and the
maximum extinction lies above the galactic plane towards galactic center and
below the galactic plane towards anti-center direction. The mean thickness of
the reddening material in the galactic plane, which is described in terms of
the half-width value $\beta$, is found to be $125\pm21$ pc. An
analytical expression for the interstellar extinction as a function of
galactic longitude and distance has been derived. The reddening analysis has
been used to constrain the Solar offset from the galactic plane. We found
that the galactic plane defined by the reddening material is symmetric around
the $z = -22.8\pm3.3$. It is seen that the galactic plane defined by the
distribution of reddening material is inclined by an angle of
$0^\circ.6 \pm 0^\circ.4$ to the formal galactic plane and the inclination is
maximum at $l \sim 54^\circ$. We summarize main results of our study in Table 3.
\begin{table}
\caption{Summary of the parameters derived in the present study}
\begin{tabular}{lcl}
\hline
Parameters           &       Value          \\ \hline
Distance of the Sun above the reddening plane  & $22.8\pm3.3$ pc   \\
Direction of maximum absorption & $l \sim 48^\circ\pm4^\circ$      \\
Angle of inclination between reddening plane & $0^\circ.6\pm 0^\circ.4$   \\
and formal galactic plane &                                  \\
Direction of maximum inclination       & $l \sim 54^\circ\pm6^\circ$      \\
Half-width value of the reddening material&   125$\pm$21 pc          \\
Scale height of the distribution of open clusters  &   53$\pm$5 pc     \\
Scale height of the distribution of reddening material  &186$\pm$25 pc \\
\hline
\end{tabular}
\end{table}
Though most of our results are in
close agreement with the values determined in previous studies but it can be
stated that a larger sample of clusters with better knowledge of their
parameters enable us to achieve  a considerable improvement in the analysis
of galactic structure. Furthermore, we provided an overview of the distribution
of reddening material in the galactic plane.

\section*{Acknowledgments}
\indent This study use the catalog given by W. S. Dias. I would like
to thank H. M. Antia and D. K.  Ojha for useful discussions during the course
of this work. The comments of the referee, Eric Schlegel, are gratefully
acknowledged.


\begin{thebibliography}{99}
\bibitem[\protect\citeauthoryear{Arce}{1999}]{Arce} Arce H. G., Goodman A. A., 1999, ApJ, 512, L135
\bibitem[\protect\citeauthoryear{Arenou}{1992}]{Arenou} Arenou F., Grenon M., G\'{o}mez A., 1992, A\&A, 258, 104
\bibitem[\protect\citeauthoryear{Bica}{2001}]{Bica} Bica E., Santiago B. X., Dutra C. M., Dottori H., de Oliveira M. R., Pavani D., 2001, A\&A, 366, 827
\bibitem[\protect\citeauthoryear{Binney}{1997}]{Binney} Binney J., Gerhard O., Spergel D., 1997, MNRAS, 288, 365
\bibitem[\protect\citeauthoryear{Cambresy}{2005}]{Cambresy} Cambr\'{e}sy L., Jarrett T. H., Beichman C. A., 2005, A\&A, 435, 131
\bibitem[\protect\citeauthoryear{Cambresy}{2001}]{Cambresy} Cambr\'{e}sy L., Boulanger F., Lagache G., Stepnik B., 2001, A\&A, 375, 999
\bibitem[\protect\citeauthoryear{Chen}{1998}]{Chen} Chen B., Vergely J. L., Valette B., Carraro G, 1998, A\&A, 336, 137
\bibitem[\protect\citeauthoryear{Chen}{1999}]{Chen} Chen B., Figueras F., Torra J., Jordi C., Luri X., Galad\'{i}-Enr\'{i}quez D., 1999, A\&A, 352, 459
\bibitem[\protect\citeauthoryear{Cohen}{1977}]{Cohen} Cohen R. S. \& Thaddeus P., 1977, ApJL, 217, L155 
\bibitem[\protect\citeauthoryear{Conti}{1990}]{Conti} Conti P. S., Vacca W. D., 1990, AJ, 100, 431
\bibitem[\protect\citeauthoryear{Dias}{2002}]{Dias} Dias W. S., Alessi B. S., Moitinho A., L\'{e}pine J. R. D., 2002, A\&A, 389, 871
\bibitem[\protect\citeauthoryear{Dutra}{2002}]{Dutra} Dutra C. M., Santiago B. X., Bica E., 2002, A\&A, 381, 219 
\bibitem[\protect\citeauthoryear{Dutra}{2003}]{Dutra} Dutra C. M., Ahumada A. V., Clari\'{a} J. J., Bica E., Barbuy B., 2003a, A\&A, 408, 287
\bibitem[\protect\citeauthoryear{Dutra}{2003}]{Dutra} Dutra C. M., Santiago B. X., Bica E., Barbuy B., 2003b, MNRAS, 338, 253
\bibitem[\protect\citeauthoryear{Fernie}{1968}]{Fernie} Fernie J. D., 1968, AJ, 73, 995
\bibitem[\protect\citeauthoryear{FitzGerald}{1968}]{FitzGerald} FitzGerald M. P., 1968, AJ, 73, 983
\bibitem[\protect\citeauthoryear{Guarinos}{1991}]{Guarinos} Guarinos J., 1991, Ph.D. Thesis, CDS, University of Strasbourg, France.
\bibitem[\protect\citeauthoryear{Gum}{1960}]{Gum} Gum C. S., Kerr F. J., Westerhout G., 1960, MNRAS, 121, 132
\bibitem[\protect\citeauthoryear{Humphreys}{1995}]{Humphreys} Humphreys R. M., Larsen J. A., 1995, AJ, 110, 2183
\bibitem[\protect\citeauthoryear{Janes}{1982}]{Janes} Janes K., Adler D., 1982, ApJS, 49, 425
\bibitem[\protect\citeauthoryear{Janes}{1988}]{Janes} Janes K. A., Tilley C., Lyng\.{a} G., 1988, AJ, 95, 771 
\bibitem[\protect\citeauthoryear{Lockman}{1977}]{Lockman} Lockman F. J., 1977, AJ, 82, 408
\bibitem[\protect\citeauthoryear{Lynga}{1982}]{Lynga} Lyng\.{a} G., 1982, A\&A, 109, 213
\bibitem[\protect\citeauthoryear{Lynga}{1985}]{Lynga} Lyng\.{a} G., 1985, in The Milky Way Galaxy, IAU Symp. 106, 133
\bibitem[\protect\citeauthoryear{Lynga}{1987}]{Lynga} Lyng\.{a} G., 1987, Computer Based Catalogue of Open Cluster Data, 5th ed. (Strasbourg CDS)
\bibitem[\protect\citeauthoryear{Magnani}{1985}]{Magnani} Magnani L., Blitz L., Mundy L., 1985, ApJ, 295, 402
\bibitem[\protect\citeauthoryear{Maiz-Apellaniz}{2001}]{Maiz-Apellaniz} Ma\'{i}z-Apell\'{a}niz J., 2001, AJ, 121, 2737
\bibitem[\protect\citeauthoryear{Martin}{1995}]{Martin} Martin C., 1995, ApJ, 444, 874
\bibitem[\protect\citeauthoryear{Mendez}{1998}]{Mendez} M\'{e}ndez R. A., van Altena, W. F., 1998, A\&A, 330, 910
\bibitem[\protect\citeauthoryear{Mermilliod}{1995}]{Mermilliod} Mermilliod J. C., 1995, in Information and On-line Data in Astronomy, ed. D. Egret \& M. A. Albrecht (Dordrecht: Kluwer), 127
\bibitem[\protect\citeauthoryear{Neckel}{1966}]{Neckel} Neckel T., 1966, Z. Astrophys., 63, 221
\bibitem[\protect\citeauthoryear{Neckel}{1980}]{Neckel} Neckel T., Klare G., 1980, AAS, 42, 251
\bibitem[\protect\citeauthoryear{Pandey}{1987}]{Pandey} Pandey A. K., Mahra H. S., 1987, MNRAS, 226, 635 (PM87)
\bibitem[\protect\citeauthoryear{Pandey}{1988}]{Pandey} Pandey A. K, Bhatt, B. C., Mahra, H. S., 1988, A\&A, 189, 66
\bibitem[\protect\citeauthoryear{Reed}{1997}]{Reed} Reed B. C., 1997, PASP, 109, 1145
\bibitem[\protect\citeauthoryear{Schlegel}{1998}]{Schlegel} Schlegel D. J., Finkbeiner D. P., Davis M., 1998, ApJ, 500, 525
\bibitem[\protect\citeauthoryear{Stenholm}{1975}]{Stenholm} Stenholm B., 1975, A\&A, 39, 307
\bibitem[\protect\citeauthoryear{Tarantola}{1982}]{Tarantola} Tarantola A., Valette B., 1982, RvGSP, 20, 219
\end{thebibliography}
\end{document}